# RECENT PROGRESS IN THE SEARCH FOR BLACK HOLES IN GALACTIC NUCLEI[1]

ROELAND P. VAN DER MAREL[2]
*Institute for Advanced Study*
*Olden Lane, Princeton, NJ 08540, USA*

**Abstract.** Massive nuclear black holes (BHs) of $10^6 - 10^9 M_\odot$ are believed to be responsible for the the energy production in quasars and active galaxies, and are thought to be present in many quiescent galaxies as well. Dynamical evidence for this can be sought by studying the dynamics of gas and stars in galactic nuclei at high spatial resolution. This paper reviews the current evidence, with emphasis on some recent developments and ongoing projects. The evidence from water masers and gas kinematics in the active galaxies NGC 4258 and M87 is compelling. In quiescent galaxies only stellar kinematics are generally available. One well-studied case is M32. Stellar dynamical $f(E, L_z)$ models with a few million solar mass BH fit the ground-based kinematical data remarkably well. N-body simulations of an edge-on $f(E, L_z)$ model for M32 show that this model is stable. HST spectra should soon provide new and improved constraints on the presence of BHs in quiescent galaxies.

## 1. Active Galaxies: Gas Kinematics

Rapid gas motions have long been known to exist in the narrow and broad emission line regions of active galactic nuclei. To use these to constrain the gravitational potential one must be able to spatially resolve the gas morphology and determine whether the motions are gravitational (or due to inflow, outflow, turbulence, etc.). This has proved very difficult with

---

astro-ph/9509055  9 Sep 1995

ground-based optical techniques, but space based and radio observations have recently yielded important progress.

In the 1970's M87 was the first galaxy for which stellar kinematical data hinted at the presence of a nuclear BH, but unambiguous interpretation remained difficult due to the unknown velocity dispersion anisotropy. The existence of rapid gas motions near the nucleus of M87 was pointed out by van der Marel (1994), who argued that these motions are gravitational. Ford et al. (1994) spatially resolved the gas morphology with HST, and showed the gas to be in a nuclear disk. Harms et al. (1994) measured gas rotation velocities in the disk of $\pm 500\,\mathrm{km\,s^{-1}}$ at $R = \pm 0.3''$, implying a nuclear dark mass of $\sim 2.4 \times 10^9 M_\odot$, probably in a BH.

Even more impressive are the VLBA radio observations of water masers in the nucleus of the active galaxy NGC 4258 (Miyoshi et al. 1995). The individual maser sources reside in a nearly edge-on torus with inner and outer radii of 4 and 8 milliarcsec. The rotation curve of the sources is Keplerian to 1% accuracy and implies a mass within the torus of $3.6 \times 10^7 M_\odot$, most likely in a BH. The high spatial resolution and small deviations from Keplerian rotation rule out most alternatives (Maoz 1995).

NGC 4258 might well remain a unique case for several years to come: only a handful of (active) galaxies has nuclear water masers, and a torus (or disk) will only maser towards the observer if seen nearly edge on. The chances of finding more evidence for BHs from the kinematics of nuclear ionized gas disks appear more promising. Several groups are now taking HST data to this extent. One interesting galaxy is the E4 radio LINER NGC 7052 (van den Bosch & van der Marel 1995). HST images show a beautiful nuclear disk of dust and ionized gas (van der Marel, van den Bosch & de Zeeuw 1996, in preparation). Ground based spectra in $0.6''$ seeing show rapid gas rotation ($250\,\mathrm{km\,s^{-1}}$ at $R = 1.5''$) and an increasing line width towards the nucleus (nuclear FWHM $550\,\mathrm{km\,s^{-1}}$). The latter can be due to seeing broadening of Keplerian rotation around a $5 \times 10^8 M_\odot$ BH, or to non-gravitational motions in an unresolved nuclear component. In Cycle 5 we will use HST to measure the gas rotation velocities at $R = 0.1''$ and $0.2''$, to obtain improved constraints on the mass of a possible BH.

## 2. Quiescent Galaxies: Stellar Kinematics

In quiescent galaxies only stellar kinematics are generally available. Unambiguous interpretation is complicated because the orbital structure is unknown, and difficult to derive from the data because of line-of-sight projection. Nonetheless, tentative evidence for nuclear BHs has been derived from ground-based data of a handful of nearby galaxies, most noticeably M32, M31, NGC 3115, NGC 3377, NGC 4594, and our own Galaxy (see

Kormendy & Richstone 1995 for an extensive review). Here I focus on M32.

M32 has a steep central rotation velocity gradient, a central peak in the velocity dispersion, and a central surface brightness cusp, all hinting at the presence of a nuclear BH. van der Marel et al. (1994a) obtained kinematical data along five different slit position angles in $\sim 0.8''$ FWHM seeing to determine line-of-sight kinematics and velocity profile shapes. A flattened stellar dynamical model with $f = f(E, L_z)$ and a $\sim 1.8 \times 10^6$ BH (or other compact dark mass) provides a remarkably good fit to these data (van der Marel et al. 1994b; Qian et al. 1995; Dehnen 1995). Kormendy & Bender (1995, in preparation) recently obtained a major axis spectrum with the CFHT in $0.5''$ FWHM seeing, which appears to indicate a somewhat larger ($\sim 3 \times 10^6$) BH mass. HST spectra of M32 will soon be obtained by van der Marel, Rix, de Zeeuw & White. To rule out models without a BH unambiguously requires the construction of flattened three-integral models. This has not been done yet, but several groups are working on this.

Kuijken & Dubinski (1994) found that flattened $f(E, L_z)$ models with much mean streaming can be unstable to the formation of a bar in their inner regions (see also Dehnen, this volume). To test whether $f(E, L_z)$ models for M32 are stable, we constructed an N-body realization of the edge-on model with a BH presented by Qian et al. (1995). The time-evolution of the system was studied with the 'self-consistent field' (SCF) N-body code of Hernquist & Ostriker (1992). Figure 1 summarizes the main results for a simulation with $2.5 \times 10^4$ equal mass particles (van der Marel et al. 1995). The axial ratios of the mass distribution, its radial profile, and the shape of the velocity ellipsoid are all constant with time. This indicates that the model is globally stable. Models for M32 that are not edge-on are intrinsically flatter. We are currently testing the stability of models with lower inclinations.

Simulations with $N = 10^{4-5}$ adequately represent the dynamics over the radial range that contains $\gtrsim 99\%$ of the mass, and suffice to study global stability. More particles are required to study the secular evolution close to the BH, and to study in detail the influence of a nuclear BH on the onset of a bar instability. We are currently running simulations with $N \gtrsim 10^6$ on a parallel computer, using the SCF code implementation of Hernquist et al. (1995) and Sigurdsson, Hernquist & Quinlan (1995). This allows a detailed study of the effects of the presence of a BH on the nuclear structure of a realistic stellar system such as M32.

Support for this work was provided by NASA through a Hubble Fellowship, #HF-1065.01-94A, awarded by the Space Telescope Science Institute which is operated by AURA, Inc., for NASA under contract NAS5-26555.

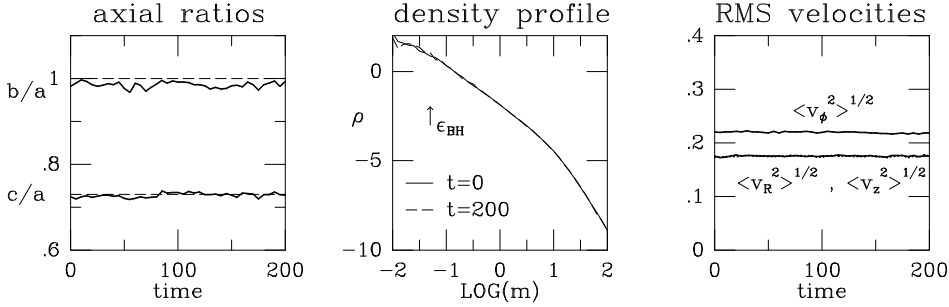

*Figure 1.* Time evolution of a self-consistent edge-on $f(E, L_z)$ model for M32 with a (softened) $1.8 \times 10^6 M_\odot$ BH, obtained from an N-body simulation with $N = 2.5 \times 10^4$. The left panel shows the 'mean' axial ratios $b/a$ and $c/a$ of the mass distribution as function of time, calculated as described in Dubinski & Carlberg (1991). Dashed lines indicate the analytical values $b/a = 1$ and $c/a = 0.73$. The middle panel shows the density profile $\rho(m)$ at the initial and final time, where $(m/a)^2 \equiv (x/a)^2 + (y/b)^2 + (z/c)^2$. The curves overlay each other. The arrow indicates the BH softening length $\epsilon_{BH} = 5 \times 10^{-2}$. The right panel shows the RMS velocities averaged over all particles, as function of time. The units of time, length and velocity used in the figures are: $1.1 \times 10^5$ yr; 30 pc = $8.7''$; and 246 km s$^{-1}$. The period of a circular orbit in the equatorial plane is $T = 0.7$ at $m = \epsilon_{BH}$ and $T = 15$ at $m = 1$. The model is globally stable, and shows no sign of a bar instability. The RMS velocities remain constant with time, so that there is probably no secular instability that introduces a dependence of the DF on a third integral. The fractional energy conservation during the simulation was $5 \times 10^{-4}$. We are currently running simulations with larger $N$ and smaller $\epsilon_{BH}$ to verify the accuracy of these results.